\documentclass[a4paper]{jpconf}
\usepackage{graphicx}
\begin{document}
\title{The Murchison Widefield Array: solar science with the low frequency SKA Precursor}

\author{S.J. Tingay{$^{1,13}$}, D. Oberoi{$^{2}$}, I. Cairns{$^{3}$}, A. Donea{$^{4}$}, R. Duffin{$^{1}$}, W. Arcus{$^1$}, G. Bernardi{$^5$}, J.D. Bowman{$^6$}, F. Briggs{$^{7,12}$}, J.D. Bunton{$^8$}, R.J. Cappallo{$^9$}, B.E. Corey{$^9$}, A. Deshpande{$^10$}, L. deSouza{$^8$}, D. Emrich{$^1$}, B.M. Gaensler{$^{11,12}$}, Goeke, R.{$^{13}$}, L.J. Greenhill{$^5$}, B.J. Hazelton{$^{14}$}, D. Herne{$^1$}, J.N. Hewitt{$^{13}$}, M. Johnston-Hollitt{$^{15}$}, D.L. Kaplan{$^{16}$}, J.C. Kasper{$^5$}, J.A. Kennewell{$^{1}$}, B.B. Kincaid{$^9$}, R. Koenig{$^8$}, E. Kratzenberg{$^9$}, C.J. Lonsdale{$^9$}, M.J. Lynch{$^1$}, S.R. McWhirter{$^9$}, D.A. Mitchell{$^{12,17}$}, M.F. Morales{$^{14}$}, E. Morgan{$^{13}$}, S.M. Ord{$^1$}, J. Pathikulangara{$^8$}, T. Prabu{$^{10}$}, R.A Remillard{$^{13}$}, A.E.E. Rogers{$^9$}, A. Roshi{$^10$}, J.E. Salah{$^9$}, R.J. Sault{$^{17}$}, N. Udaya-Shankar{$^{10}$}, K.S. Srivani{$^{10}$}, J. Stevens{$^8$}, R. Subrahmanyan{$^{10,12}$}, M. Waterson{$^1$}, R.B. Wayth{$^{1,12}$}, R.L. Webster{$^{17,12}$}, A.R. Whitney{$^9$}, A. Williams{$^{18}$}, C.L. Williams{$^{13}$} and J.S.B. Wyithe{$^{17,12}$}}

\address{$^{1}$ ICRAR - Curtin University, Perth, Australia}
\address{$^{2}$ National Centre for Radio Astrophysics, Pune, India}
\address{$^{3}$ The University of Sydney, Sydney, Australia}
\address{$^{4}$ Monash University, Melbourne, Australia}
\address{$^{5}$ Harvard-Smithsonian Center for Astrophysics, Cambridge, MA, USA}
\address{$^{6}$ Arizona State University, Tempe, AZ, USA}
\address{$^{7}$ The Australian National University, Canberra, Australia}
\address{$^{8}$ CSIRO Astronomy and Space Science, Australia}
\address{$^{9}$ MIT Haystack Observatory, Westford, MA, USA}
\address{$^{10}$ Raman Research Institute, Bangalore, India}
\address{$^{11}$ Sydney Institute for Astronomy, The University of Sydney, Sydney, Australia}
\address{$^{12}$ ARC Centre of Excellence for All-sky Astrophysics (CAASTRO)}
\address{$^{13}$ MIT Kavli Institute for Astrophysics and Space Research, Cambridge, MA, USA}
\address{$^{14}$ University of Washington, Seattle, USA}
\address{$^{15}$ School of Chemical and Physical Sciences, Victoria University of Wellington, New Zealand}
\address{$^{16}$ University of Wisconsin--Milwaukee, Milwaukee, WI, USA}
\address{$^{17}$ The University of Melbourne, Melbourne, Australia}
\address{$^{18}$ ICRAR - University of Western Australia, Perth, Australia}

\ead{s.tingay@curtin.edu.au}

\begin{abstract}
The Murchison Widefield Array is a low frequency (80 - 300 MHz) SKA Precursor, comprising 128 aperture array elements (known as tiles) distributed over an area of 3 km diameter.  The MWA is located at the extraordinarily radio quiet Murchison Radioastronomy Observatory in the mid-west of Western Australia, the selected home for the Phase 1 and Phase 2 SKA low frequency arrays.  The MWA science goals include: 1) detection of fluctuations in the brightness temperature of the diffuse redshifted 21 cm line of neutral hydrogen from the epoch of reionisation; 2) studies of Galactic and extragalactic processes based on deep, confusion-limited surveys of the full sky visible to the array; 3)  time domain astrophysics through exploration of the variable radio sky; and 4) solar imaging and characterisation of the heliosphere and ionosphere via propagation effects on background radio source emission.  This paper concentrates on the capabilities of the MWA for solar science and summarises some of the solar science results to date, in advance of the initial operation of the final instrument in 2013.
\end{abstract}

\section{Introduction}
The Murchison Widefield Array (MWA) is the only Square Kilometre Array (SKA: \cite{ska}) Precursor telescope at low radio frequencies.  An SKA Precursor is a recognised SKA technology demonstrator located at one of the two sites that will host the SKA, the Murchison Radio-astronomy Observatory (MRO) in the Murchison region of Western Australia and the Karoo region of South Africa's Northern Cape.  The MWA is sited at the MRO, along with a second SKA Precursor, the Australian SKA Pathfinder (ASKAP: \cite{jon08}; \cite{jon07}).  The MeerKAT SKA Precursor will be located at the South African site\footnote{http://public.ska.ac.za/meerkat}.  The MWA is sited at the MRO due to the extremely low levels of human-made radio frequency intereference in this area of Western Australia, particularly within the FM band (87 - 108 MHz) which is encompassed by the MWA at the low end of its operating frequency range.

The technical capabilities of the MWA are described in detail by \cite{tin12} and the full science case for the MWA is described by \cite{bow12}.  Broadly, the MWA science goals include: 1) detection of fluctuations in the brightness temperature of the diffuse redshifted 21 cm line of neutral hydrogen from the epoch of reionisation; 2) studies of Galactic and extragalactic processes based on deep, confusion-limited surveys of the full sky visible to the array; 3)  time domain astrophysics through exploration of the variable radio sky; and 4) solar imaging and characterisation of the heliosphere and ionosphere via propagation effects on background radio source emission.  

The MWA has reached the point of practical completion and will be available to the global user community from mid-2013, based on an ``Open Skies'' policy (http://www.mwatelescope.org).  Users will be able to propose observational programs with the MWA in the first half of 2013, including for solar science.  Considerable data have been obtained in advance of completion of the full instrument.  A prototype array was operated for approximately 2 years, from 2009 to 2011, and during construction of the full array, science commissioning observations were undertaken.

This paper will briefly describe the capabilities of the MWA for solar physics and summarise some of the solar results thus far obtained.  The results to date have been obtained with arrays four times less sensitive than the final array that will be available in 2013, and generally with lower angular resolution and image fidelity than will be available with the final array.

\section{History of low frequency solar imaging in Australia}
Low radio frequency imaging of the Sun was pioneered in Australia under the leadership of Paul Wild \cite{Eke12}.  Wild and others imaged the Sun at frequencies similar to the MWA frequencies for the first time, revealing the nature of type II and type III bursts (these types were defined by the same group based on high quality dynamic spectra) and monitoring their evolution as a function of time and frequency \cite{wil68}.  This work used the Culgoora solar heliograph, built by Wild and his team in the late 1960s.

Since the closure of Culgoora in 1984, there has been no low frequency solar imaging capability in Australia. The MWA will therefore renew this capability in Australia using modern instrumentation on a radio quiet site.

\section{Results from the MWA prototype and science commissioning}
Important capabilities of the MWA for solar physics are the time and frequency resolution of imaging.  The MWA can produce images at a rate of two per second, for each of the four Stokes parameters and for each frequency channel across the bandwidth.  The bandwidth is 30.72 MHz, tunable in the range 80 to 300 MHz, with the 30.72 MHz broken up into a maximum of 3072 channels of 10 kHz each.  Thus, in one minute it is possible for the MWA to produce almost 1.5 million images (number of frequency channels $\times$ number of Stokes parameters $\times$ number of time steps), allowing a comprehensive exploration of the evolution of the radio emission as function of both frequency and time.

The angular resolution of the MWA at these low frequencies is of order 1 arcminute, likely within a factor of a few of the expected scattering size due to the intervening plasma, at the top end of the MWA frequency range.  Approximately 1000 resolution elements will cover the disk of the Sun, at the top end of the MWA frequency range (fewer resolution elements at lower frequency).

In Novemnber 2012 the MWA reached a major milestone in practical completion of the full instrument as described by \cite{tin12} and in early 2013 the full capabilities of the MWA will be exercised for the first time, leading to commencement of the first science operations phase from mid-2013.  During the period 2009 - 2011, a 32 tile prototype array was operated and used as a science and engineering testbed.  This was decommissioned in early 2012, to make way for the construction of the final instrument.  In the final stages of construction, from approximately September 2012, science commissioning of the final instrument commenced, overlapping with construction; 32 tile subsets of the final instrument were commissioned and tested as they were completed.  The 32 tile sub-arrays ranged from configurations with only short baselines (low angular resolution) to configurations with close to the maximum length baselines (high angular resolution).

During both the prototype phase and science commissioning phase, scientific observations were performed across all MWA science themes.  In the prototype phase, \cite{obe11} performed the first demonstration of the spectroscopic imaging capabilities of the MWA (albeit with relatively low angular resolution), revealing a rich variety of phenomena in the observed frequency range of 171 - 202 MHz, consisting of many short-lived, narrow-band, non-thermal emission features.  While the \cite{obe11} images are of low angular resolution, they show a number of these events to be marginally resolved.  A somewhat surprising feature of these first serious solar observations with the MWA was the rich variety of weak transient events, showing that to effectively study the Sun at these frequencies a surperb radio quiet location is required.  Even to study the Sun, one of the brightest radio objects in the sky, very low levels of human-made interference appear critical to success.

Other MWA prototype observations of Type III flares from 2011 are currently being analysed in conjunction with SDO and RHESSI data (Cairns et al. 2013, in preparation).  Even the prototype phase of the MWA development was scientifically productive in terms of solar science.

In the science commissioning phase of the final MWA instrument, since approximately September 2012, almost 100 TB of solar imaging data have been collected!  This vast store of data covers observations with five different 32 tile sub-arrays, generally during quiet periods for the Sun.  The most effective sub-array was the so-called ``gamma'' array, since the configuration of 32 tiles used included long baselines and therefore produced high angular resolution images (few arcminute angular resolution).  An example is shown in Figure 1, from data obtained on October 29, 2012 (03:35:44 UT).  This image represents a single 1 second integration and a 40 kHz channel at a frequency of approximately 150 MHz in a single linear polarisation.  The disk of the Sun is well-resolved.  The Sun on this day was rated as quiet.  However, even in this state weak transient events are clearly seen when a series of images are made over a period of several minutes (Tingay et al. 2013, in preparation).

A limb brightening effect could be expected for the data presented in Figure 1 \cite{kun64}.  However, with an angular resolution of only $\sim$4 arcminutes ($>$10\% of the solar disk diameter), the brightening is likely to be smeared into the disk emission.  At 150 MHz, the altitude of the resonant plasma lies at a distance of around 1.2 Rs, which gives a diameter of around 38 arcminutes at the Earth.  This is approximately the diameter of the Sun seen Figure 1.  We take this as evidence for additional limb emission due to the corona and the absence of an apparent change in the surface brightness at the limb because of the smearing due to limited angular resolution.

\begin{figure}[h]
\centering
\includegraphics[width=15cm]{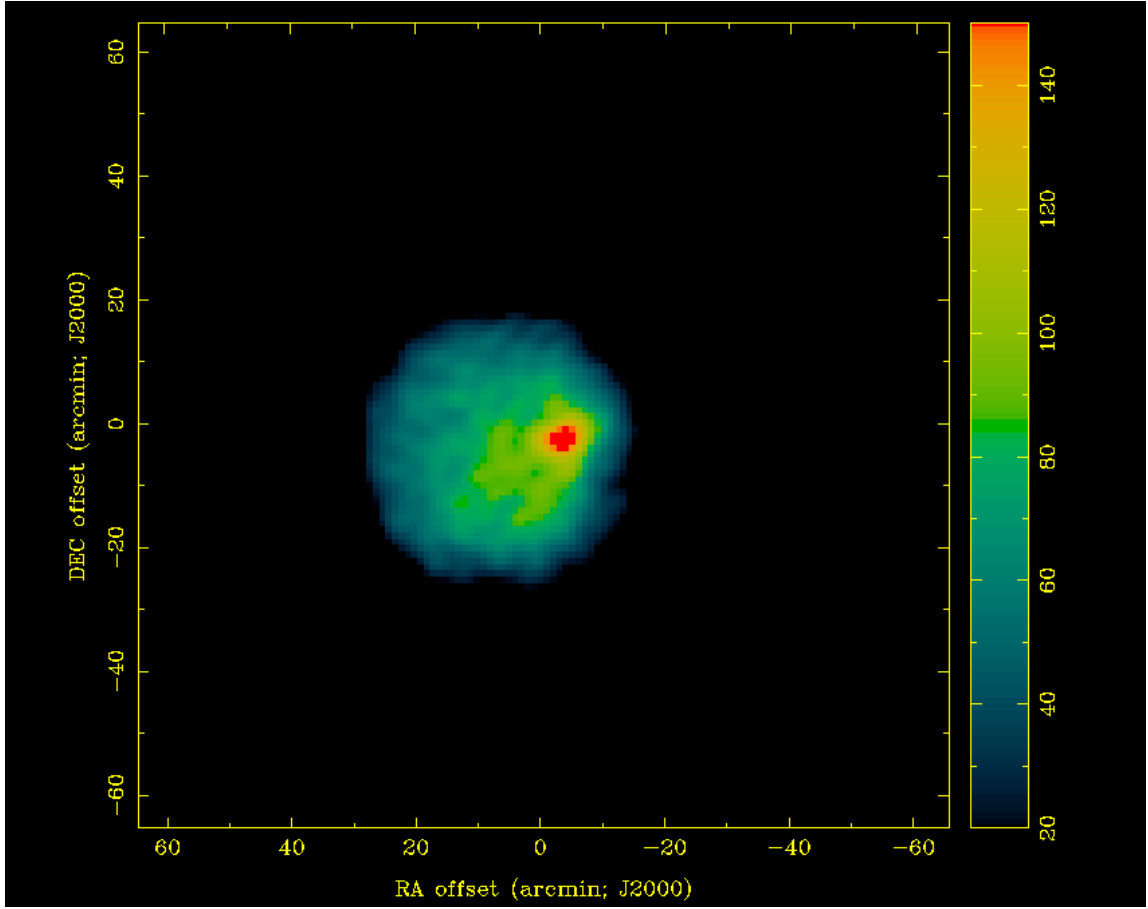}
\caption{An image from the science commissioning phase of the MWA.  Data obtained as described in the text.  Normal astronomical units are used, with north at top and east to the left.  The synthesised beam has a full width at half maximum of approximately 4 arcminutes and is close to circular.}
\end{figure}

\section{Future prospects}
In addition to the capabilities of the final instrument described above, the MWA will have the ability to record the raw voltage streams from each tile.  This capability makes it possible to produce dynamic spectra and images at very high time and frequency resolution, sub-millisecond and sub-10 kHz (but not at maximum time and frequency resolution simultaneously).  Such high resolution imaging will open up the ability to study the temporal and frequency sub-structure of both strong flares and the weak features seen by \cite{obe11}.

The MWA data will be of most use if they can be combined with data from other solar observation platforms in other wavelength ranges.  Part of the motivation for attending this conference and providing an update on the capabilities of the MWA was to make contact with other instrument teams and to start to coordinate MWA solar observations with multi-wavelength campaigns.

As described in \cite{bow12}, the MWA could play a significant role in tracking large Coronal Mass Ejections and making early predictions regarding collisions with the Earth.  It is thus timely that the MWA will come into full operation in the first half of 2013, presumably the year of maximum activity during Cycle 24.

\section*{Acknowledgments}
This scientific work makes use of the Murchison Radio-astronomy Observatory. We acknowledge the Wajarri Yamatji people as the traditional owners of the Observatory site. Support for the MWA comes from the U.S. National Science Foundation (grants AST-0457585, PHY-0835713, CAREER-0847753, and AST-0908884), the Australian Research Council (LIEF grants LE0775621 and LE0882938), the U.S. Air Force Office of Scientic Research (grant FA9550-0510247), and the Centre for All-sky Astrophysics (an Australian Research Council Centre of Excellence funded by grant CE110001020). Support is also provided by the Smithsonian Astrophysical Observatory, the MIT School of Science, the Raman Research Institute, the Australian National University, and the Victoria University of Wellington (via grant MED-E1799 from the New Zealand Ministry of Economic Development and an IBM Shared University Research Grant). The Australian Federal government provides additional support via the National Collaborative Research Infrastructure Strategy, Education Investment Fund, and the Australia India Strategic Research Fund, and Astronomy Australia Limited, under contract to Curtin University. We acknowledge the iVEC Petabyte Data Store, the Initiative in Innovative Computing and the CUDA Center for Excellence sponsored by NVIDIA at Harvard University, and the International Centre for Radio Astronomy Research (ICRAR), a Joint Venture of Curtin University and The University of Western Australia, funded by the Western Australian State government.

\end{document}